\documentclass[english,aps,pra,reprint,noshowpacs,superscriptaddress,nofootinbib]{revtex4-1}   
\usepackage[T1]{fontenc}	
\usepackage[latin9]{inputenc}	
\usepackage{geometry}		
\usepackage{graphicx}
\usepackage[above,below]{placeins}	
\usepackage{times}

\usepackage{hyperref}
\usepackage{array}
\hypersetup{colorlinks=true,urlcolor=blue,citecolor=blue,linkcolor=blue}   
\begin{document}

\title{A controlled study of stereoscopic virtual reality in freshman electrostatics}
\author{J. R. Smith}
\author{A. Byrum}
\author{T. M. McCormick}
\author{N. Young}
\author{C. Orban}
\author{C. D. Porter}
\affiliation{Physics Department, The Ohio State University, 191 W. Woodruff Ave., Columbus, Ohio, 43210}


\begin{abstract}
Virtual reality (VR) has long promised to revolutionize education, but with little follow-through. Part of the reason for this is the prohibitive cost of immersive VR headsets or caves. This has changed with the advent of smartphone-based VR (along the lines of Google cardboard) which allows students to use smartphones and inexpensive plastic or cardboard viewers to enjoy stereoscopic VR simulations. We have completed the largest-ever such study on 627 students enrolled in calculus-based freshman physics at The Ohio State University. This initial study focused on student understanding of electric fields. Students were split into three treatments groups: VR, video, and static 2D images. Students were asked questions before, during, and after treatment. Here we present a preliminary analysis including overall post-pre improvement among the treatment groups, dependence of improvement on gender, and previous video game experience. Results on select questions are discussed.

\end{abstract}

\maketitle

\section{Introduction}

\label{sec_intro} 
Student success in STEM curricula often hinges on the ability of an instructor to portray physical phenomena and mathematical representations to a classroom~\cite{discipline}. Particularly in STEM disciplines, students are faced with a variety of concepts and systems that are inherently three-dimensional (3D) in nature and involve either microscopic or invisible phenomena. The need for useful visualizations is clear and has been met with a variety of innovative projects such as PhET interactives, videos and simulations. However, all of these visualization methods have one thing in common: they visualize the 3D world using 2D media. 

Virtual reality (VR) technology has advanced rapidly in recent years, and the past year has seen a proliferation of VR in consumer products, education, professional training, and public outreach \cite{NASA}.
Meanwhile, smartphone technology has reached the point where 3D animations can be rendered in real time on most devices. Realizing that smartphone technology can provide an inexpensive method for producing VR, in 2014 two Google engineers created a small headset called ``Google Cardboard'' to do precisely this~\cite{Google}. 
With an investment of a few dollars for cardboard headsets, this enables instructors to regularly incorporate VR into the classroom environment as a sustained learning tool. This technology can make complex 3D systems easier to visualize, and can provide a much more immersive visualization experience than 3D animated videos. It also lifts certain constraints on what questions can conveniently be asked by instructors; assessments are no longer limited to systems with convenient 2D projections.

Although it is beyond the scope of this work to comprehensively review the literature, there have been dozens of investigations into the use of VR in the classroom. Examples include work on virtual environments using a 2D computer screen \cite{osuVR}, VR headsets \cite{Kauf}, and immersive stereoscopic views in VR ``caves''~\cite{John}. Although there are exceptions this rule, many prior studies have not had control groups and have had low numbers of participants, due in part to sharing of a single, expensive VR device. In order to investigate the utility of this technology in introductory physics, smartphone apps were written by the investigators to illustrate several 3D systems and concepts from first-year electrostatics. This paper provides a preliminary report on efforts related to electric fields around charged particles, especially dipole fields. This discussion will include a description of the apps and the engine used to produce them, the questions asked of students to assess their learning and attitudes, and salient findings about the utility of VR over alternatives (in this case, video and static 2D images).

\section{Methods}
{
\label{sec_methods}

Throughout the spring semester of 2017, assessments were given to students in the second semester of OSU's calculus-based introductory physics course. 
All interaction with students occurred after initial instruction on electric fields in lecture, recitation and lab had concluded. Students were asked a series of computer-based multiple-choice questions related to electric fields around charged particles. 
These questions were split into two parts: one before any treatment and one after a treatment. Students were given points equivalent to one homework assignment for completing the entire exercise; their points were not dependent on the correctness of their answers. Students were given the opportunity to opt out of research participation and the data from those who did opt out have been removed. 
89\% of the 707 students contacted at the beginning of the semester participated in some part of the VR study. Of these, 301 were assigned to the electric field portion, which is relevant to this paper. 210 of these students reported their sex as male, 66 as female, and 25 declined to provide a sex. 

\subsection{Questions}

Students were asked basic questions about the direction of the electric field at various points in space around one or more charged particles. Representative examples of the questions asked are shown in Fig.~\ref{fig_qs}. As indicated in the figure, some question stems included 2D (\textit{xy}) axes, and others included 3D (\textit{xyz}) axes. This was done with the intent of determining whether gains from any treatment type were robust to presentation style or against the priming of the axes given. Questions on the pre- and post-test portions were very similar, but were identical in only a few cases. A typical example of this minor difference between pre-post questions is shown in Fig.~\ref{fig_qs}(a) and (b), which both ask students about the electric field along the axis of the dipole, but not between the two charges. For more on the types of questions asked, see Sec.\ \ref{sec_results}. Students were asked 33 questions about electric field directions, 3 attitudinal questions (whether they enjoyed the treatment, found it helpful, and would recommend it to a friend), and 2 questions on the frequency with which they currently play video games and the frequency they played as adolescents.

\begin{figure}
\includegraphics[width=0.8\linewidth]{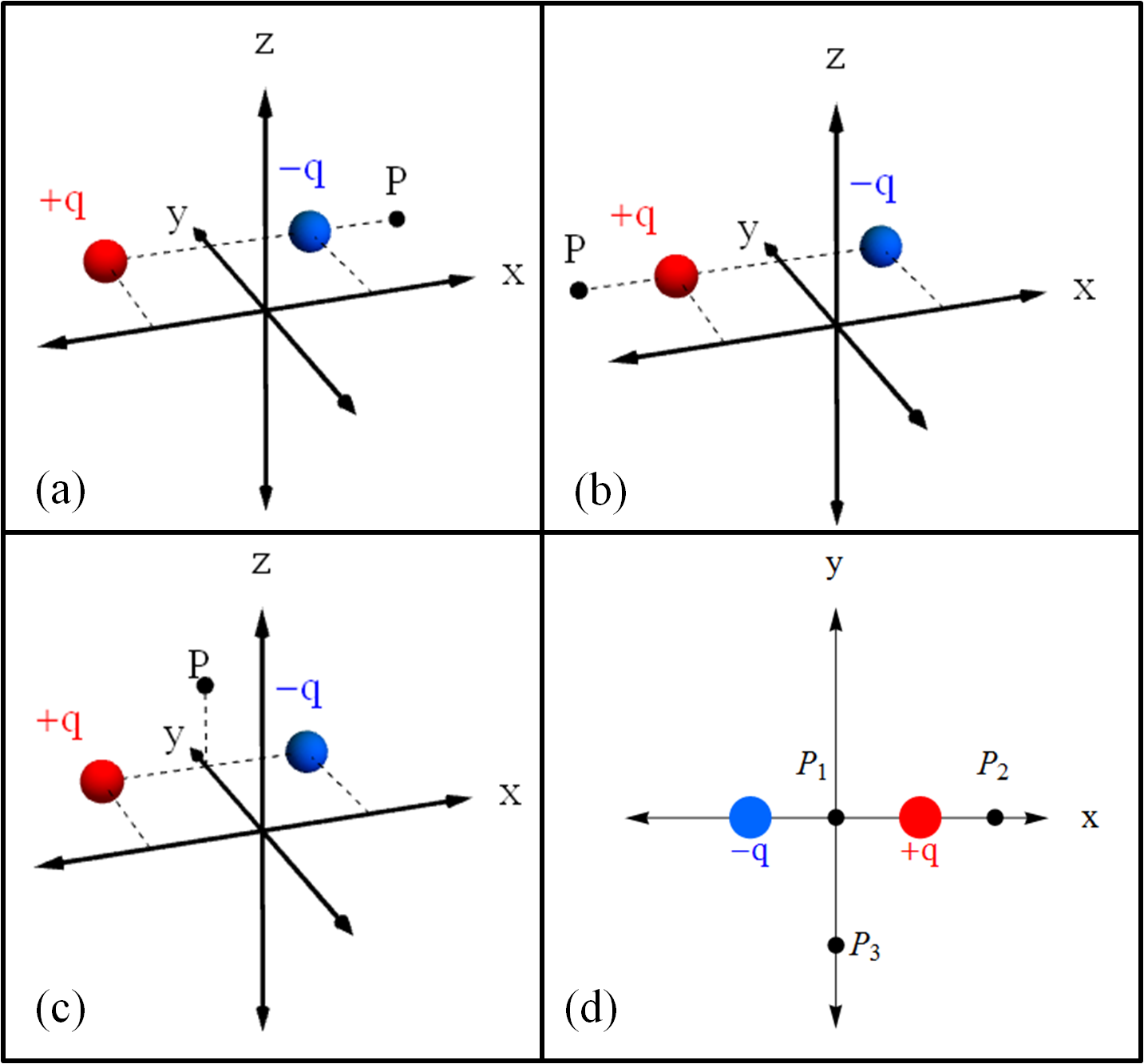}
\caption{Students were always asked to select from multiple choice options, the direction of the electric field at a point labeled $P$ or $P_{1}$, etc. Here are four examples of system diagrams and labeled points. \label{fig_qs}}
\end{figure}


\subsection{Treatments}

Students were scheduled to take the assessments in groups. As students entered, they were randomly assigned to one of three treatment types (virtual reality, video, and static images). 
The assessments were identical for all students, only the type of training between the pre- and post-test portions differed. All treatment types involved visualizations of electric dipoles and the fields surrounding them. Post-hoc analysis showed that students' final scores in their physics course were comparable between treatment groups (VR: $78\%$, Video: $75\%$, Images: $76\%$).


\textbf{Virtual reality:} 
VR visualizations were created as applications for Android smartphones. The apps were written using Unity, a cross-platform game engine developed by Unity Technologies \cite{Unity}, primarily used in development of video games and simulations for gaming consoles, computers and mobile devices, and the Google VR SDK for Unity. For the application used in this portion of the VR study, two oppositely-charged particles were modeled as point charges and visually represented as spheres as illustrated in Fig.~\ref{fig_dip}. The electric field was represented as an array of vectors generated in Unity, where the direction and magnitude of each vector was determined by the distance from charges in the system. 
The application was then built as an Android application package (APK file), and installed on two Nexus-5 smartphones.\footnote{The APK file used in this study is available by request. Several electric field visualizations, including a version of the dipole visualization used here, are publicly available on Google Play~\href{http://go.osu.edu/BuckeyeVR}{go.osu.edu/BuckeyeVR}.} 

The app splits the phone into two screens, one for each eye. 
The phone is then placed in a \$2.50 cardboard viewer. The students can now view the electric dipole depicted in Fig.~\ref{fig_dip} in stereoscopic 3D.
The app utilizes the smartphone's sensors so that when the students turn their heads, the dipole to rotates so that they can see it from any orientation. Students were given instructions to ``look around'' and study the electric field vectors from many angles before returning to the assessments. The VR sequence consisted of five instruction-visualization pairs, where the screen displayed instructions for 15-seconds and then allowed students to control the visualization for 25-seconds, for a total of a 3-minute, 20-second treatment time. 

\begin{figure}
\includegraphics[width=0.8\linewidth]{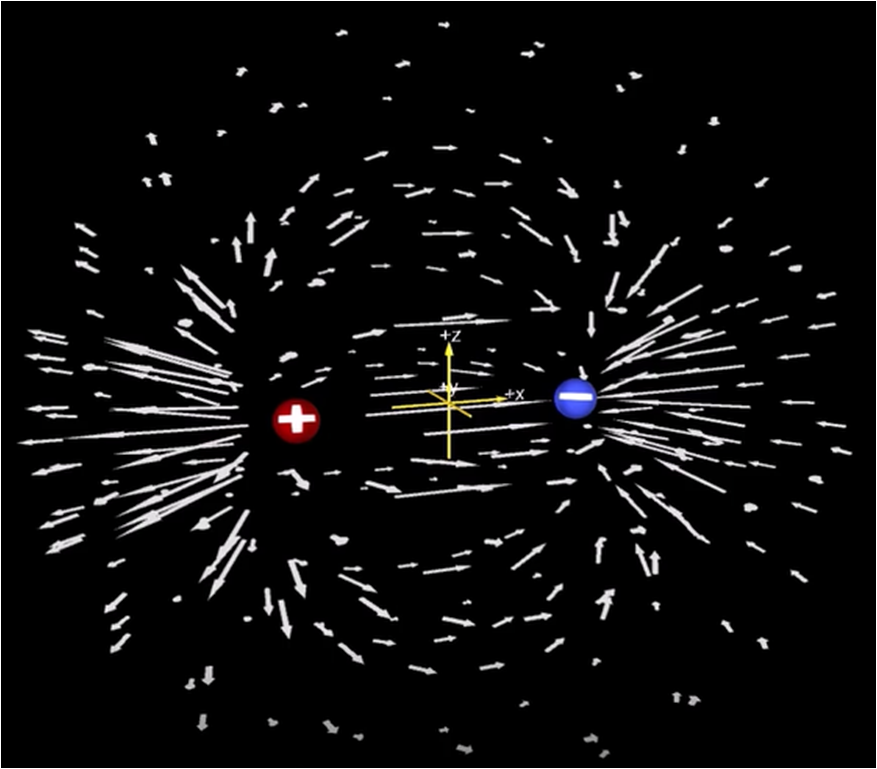}
\caption{An electric dipole with electric field vectors generated at select points around the charges. This image is indicative of the style and detail seen by all treatment groups. Several electric field visualizations, including a version of the dipole visualization shown above, are publicly available on Google  Play~\href{http://go.osu.edu/BuckeyeVR}{go.osu.edu/BuckeyeVR}\label{fig_dip}.}
\end{figure}
}

\textbf{Videos:} A 2-minute, 50-second video was generated from the application described above. The screen was not split. Students in the ``video'' treatment group watched the video, which has intermittent instructions on what details students should be noting. Students in this treatment group saw all the same details as those in the VR group; the only difference being that those in the VR group saw it in stereoscopic 3D and were in control of their viewing of the system. 


\textbf{Images:} Students in the ``images'' treatment group were shown static 2D images of dipoles and the electric fields around them. Students were shown a static screenshot from the video/VR like the one shown in Fig.\ \ref{fig_dip}. Since these students were shown only the type of visualization found in their textbooks and shown to them in class, this "treatment'' serves as a control group.
\section{Results}
{
\label{sec_results}

Table~\ref{tab_null_hyp} presents average gains between the pre- and post-test on question pairs (with the post-test question only slightly altered, if not repeated exactly). For eight different question pairs there were no significant ($>0.05$) differences between the three treatment types. This was surprising to us as we had expected the VR treatment to significantly outperform the two other treatments.


\begin{table}
\caption{A brief description of question types that were separately considered is included in the first column. Mean post-pre gains are shown as a percent of possible points in that question type. 
The far right column shows the $p$ values for the hypothesis that the mean gains between treatment types are different. In all cases this hypothesis is rejected. 
\label{tab_null_hyp}}
\centering

\begin{tabular}{| c | c | c | c | c |}
 \hline
 Question type & \multicolumn{3}{|c|}{Post-pre gains ($\%$)} & t-test \\
 \hline
  & VR & Video & Image & $p$\\
 \hline
 $P$ on the axis of the dipole, $E \neq 0$ & 10.4 & 6.5 & 6.6 & 0.18 \\ 
 \hline
 $P$ on some axis, but not along & & & & \\ the dipole axis, $E \neq 0$. & 10.6 & 12.4 & 10.6 & 0.92 \\ 
 \hline
 $P$ off axis, but in $xy$, $yz$, or $xz$ & & & & \\ planes, $E \neq 0$. & -1.9 & 1.9 & -0.5 & 0.10 \\ 
 \hline
 $P$ off axis, out of plane and $E \neq 0$. & -6.7 & 0.2 & -4.9 & 0.72\\
 \hline
 $E = 0$. & 1.0 & 1.0 & 2.4 & 0.64\\
 \hline
 2D axes & 15.9 & 11.3 & 12.3 & 0.19 \\
 \hline
 3D axes & -9.2 & -3.9 & -6.7 & 0.17 \\
 \hline
 More than one charged particle & 3.0 & 3.8 & 3.3 & 0.74 \\
 \hline
\end{tabular}
\end{table}

There were significant differences among students' attitudes to the different treatments. Student data on their degree of enjoyment are detailed in Fig.~\ref{fig_att}. The mean ``enjoyable'' rankings (with ``highly unenjoyable'' = -2 and ``highly enjoyable'' = +2) were significantly different for those students given the VR treatment (mean = 1.1) and for those given other treatments (mean = 0.68), according to a $\chi^2$ test ($\chi^2=153$, $p<0.01$).  Using a similar scale, students reported on how helpful they perceived the treatments to be. Here also the VR treatment (mean = 1.06) scored significantly higher than the others (mean = 0.70), with $p<0.01$.
\begin{figure}
\includegraphics[width=0.85\linewidth]{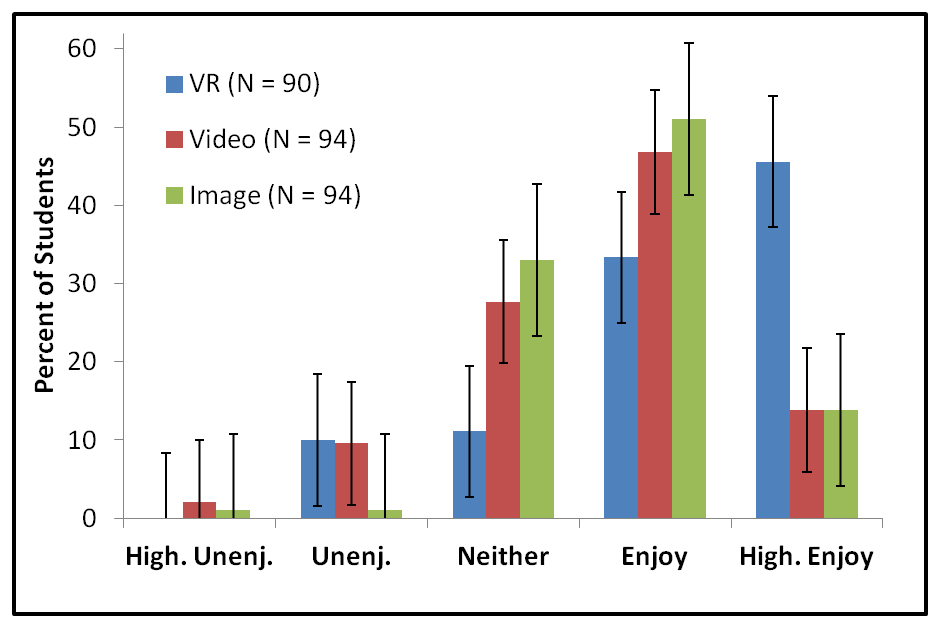}
\caption{A histogram of students' reported enjoyment of the visualizations in each treatment group. The error bars describe standard error.\label{fig_att}}
\end{figure}


Researchers have found gender differences in spatial abilities across cultures \cite{Silverman2007}. As shown in Table~\ref{gender}, there were significant differences between pre-post gains for men compared to women in the case of the VR treatment, but not for the video or image treatment, with men's gains being larger than women's gains in the VR treatment group. Some authors propose video game training as a means to ameliorate gender gaps in spatial reasoning skills~\cite{Spence}. With this in mind, our study asked participants about the frequency of their video game play both current and prior. Frequent video game play is defined as at least once per week. Table~\ref{gender} shows those results for each treatment group. In all cases, men's self-reported video game play is higher than women's.

\begin{table}
\caption{Correlation coefficients and $p$ values showing the relationship between video game play, gains between pre- and post-test score, and gender. A positive correlation indicates larger values for men. \label{gender}}
\begin{center}
\begin{tabular}{| c | c | c | c | c | c | c |}
 \hline
  & \multicolumn{2}{|c|}{VR} & \multicolumn{2}{|c|}{Video} & \multicolumn{2}{|c|}{Image} \\
 \hline
  & C.C. & $p$ & C.C. & $p$ & C.C. & $p$ \\
 \hline
  Gains vs. gender & 0.25 & 0.01 & -0.02 & >0.8 & -0.15 & 0.14\\ 
  \hline
 Curr. video game & & & & & & \\
 vs. gender & 0.48 & <0.01 & 0.60 & <0.01 & 0.56 & <0.01 \\ 
 \hline
 Prior video game & & & & & &  \\ 
 vs. gender & 0.49 & <0.01 & 0.45 & <0.01 & 0.56 & <0.01 \\ 
 \hline
 
\end{tabular}
\end{center}
\end{table}

In the next analysis we performed, the data were separated according to self-reported prior video game play (regardless of gender). It was hypothesized that those students reporting greater video game play might be less distracted or overwhelmed by the 3D rotations and might get more out of the VR simulation. Of all questions asked on the assessments, there were 7 pairs that were repeated on the pre- and post-tests, and these 7 pairs were related to the electric dipole, such as images (a) and (b) of Fig.~\ref{fig_qs}. Let $S^{VR}_{\rm pre,oft}$ be the average score on a pre-treatment (pre) question by students who reported playing video games often (oft, defined as more than once per week) and received the VR treatment (VR). Score $S^{\rm vid}_{\rm post,rare}$ is the average score on a post-test question, for students who reported playing video games rarely (rare, defined as once per week or fewer), and received the video (vid) treatment, and so on. We report not simply the pre-post gains, but the difference between pre-post gains for those who report video game play often, and the pre-post gains for those who report video game play as rare. That is,
\begin{eqnarray}
\Delta G=(S_{\rm post,oft}-S_{\rm pre,oft})-(S_{\rm post,rare}-S_{\rm pre,rare}).\nonumber
\end{eqnarray}
\begin{figure}
\includegraphics[width=0.95\linewidth]{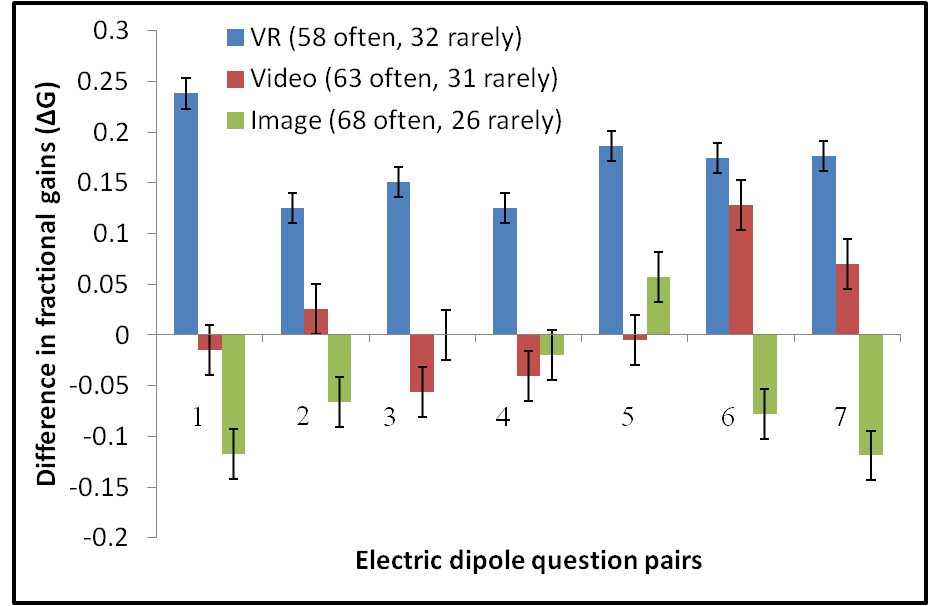}
\caption{The difference $\Delta G$ between post-pre gains for those reporting playing video games often and those reporting doing so rarely. This difference is shown for seven pairs of pre-post questions related to dipole electric fields. For example, pair 3 refers to the setups displayed in Fig.\ \ref{fig_qs} (a) and (b). The error bars describe standard error. \label{fig_gains_diff}}
\end{figure}
In all cases, the gains for students in the VR treatment group depend significantly on whether or not they report regular video game play. More specifically, among all students reporting regular video game play, those who received the VR treatment had significantly higher gains on these seven paired questions than did those who received the video treatment (t-test, $t=1.7$, $p=0.04$, $d=0.30$), and those who received the image treatment (t-test, $t=2.4$, $p<0.01$, $d=0.42$). Among those students who reported rarely playing video games, students given the VR treatment had lower gains than either of the other treatments, but these differences were not significant to the 0.05 level. These results support the hypothesis that successful use of VR is dependent on the student having some experience with video games (which we use as a proxy for 3D visuospatial rotations familiarity). The authors note here that since many correlations have been sought in these data, the chances of finding false positives at the 0.05 level are non-negligible. We intend to follow up our preliminary findings in future work.
}

\section{Conclusions and Future Work}

In conclusion, we find evidence that the VR treatment can be more effective for students with a video-gaming background. In samples of students with or without a video-gaming background, there does appear to be a slight gender bias against women for the effectiveness of VR treatments, and this could be due to the disparity between men's and women's self-reported experience with video games in our study. Students show significant preference for VR over other treatments in terms of perceived helpfulness, enjoyment, and recommendations they would make to friends.   

In future work a natural question is whether repeated exposure to VR can improve scores and ameliorate gender differences for students who do not report frequent video game use. 


\acknowledgments{The development of VR visualizations has been supported by the OSU STEAM Factory, OSU's Marion campus, and the Office for Distance Education and eLearning (ODEE).}

\bibliographystyle{unsrt}
\bibliography{main}
















\end{document}